# Concurrent Coding – A Reason to Think Differently About Encoding Against Noise, Burst Errors and Jamming

David M Benton[*]
[1]Aston University, Aston Triangle, Birmingham, B4 7ET
[*] *d.benton@aston.ac.uk*

**Abstract**
Concurrent coding is an unconventional encoding technique that simultaneously provides protection against noise, burst errors and interference. This simple-to-understand concept is investigated by distinguishing 2 types of code – open and closed with the majority of the investigation concentrating on closed codes. Concurrent coding is shown to possess an inherent method of synchronisation thus requiring no additional synchronisation signals to be added. This enables an isolated codeword transmission to be synchronised and decoded in the presence of noise and burst errors. Comparisons are made with the spread spectrum technique CDMA. With a like-for-like comparison concurrent coding performs comparably against random noise effects, performs better against burst errors and is far superior in terms of transmitted energy efficiency.

## 1. Introduction

The majority of approaches currently in use to protect against data loss or corruption, involve encoding techniques that operate independently upon statistical errors, non-random errors and interference. Correcting statistical errors might involve adding information into the data stream about the structure of the data. This is typically inserted locally close to the data and helps to overcome the effects of noise. Non-random errors, such as burst errors that remove contiguous bits of the data, can exceed the recovery capability of the statistical error routine and are overcome using interleaving to disperse the connected information. Interference might be overcome with the use of a spread spectrum technique. Concurrent coding is a very different way of protecting information [1]-[5] which performs all of the above functions in a single process[6]. Concurrent coding represents a different approach where data is redistributed 'globally' within a data set rather than locally. The local encoding principle is widely seen in block codes [7][8] including cyclic codes, Golay codes and Reed Solomon encoding[9]. Local encoding is also found in convolutional coding schemes such as Turbo codes[10] and Viterbi codes[11]. Generally, local encoding does not deal efficiently with non-random errors [12][13]and typically interleaving would also be required. Methods such as Fire codes[14] and Reed Solomon encoding [9],[15], treat data as a set of symbols and correct for symbol errors, encompassing non-random errors. Concurrent codes are an asymmetric binary encoding system where digital 1's are represented by indelible marks placed into a codeword. Marks have a substantive presence, such as energy pulses and cannot be removed on a random basis (which would convert a 1 back to a 0). A zero is the absence of a positive signal presence or substance, but 0's can be converted to 1's by noise (or jamming).

Concurrent coding distributes the individual data words throughout the entire data set through the use of a hashing function, applied iteratively to increasing size prefixes of the data word. A variety of hashing functions have been used [4][6][16] with an emphasis on distribution rather than security although algorithm attacks have been examined[17]. Recently a new concurrent code based on the use of monotone Boolean functions has been developed by Hanifi *et al* [18].

Some salient properties of concurrent coding

- Encoded message vectors will always be decoded. - a result of using indelible marks.
- Original message vectors can be obscured by false decodings, called hallucinations, but cannot be removed or corrupted.
- Individual marks in the codeword can be shared by many input vectors (which have the same sub-sequence of bits), reducing the number of '1's and leading to improved efficiency in terms of transmitted energy.
- A significant benefit of concurrent coding in comparison with other resilient encoding techniques, such as cyclic encoding, is that it is really very simple to understand.

With current concerns about energy usage and the role that communications plays in this [19] the improved efficiency of concurrent coding is a particularly interesting aspect.

The principles of concurrent coding are laid out in previous references [1]-[6] but are given here briefly. The concurrent coding principle hashes incrementally increasing subsections (or prefixes) of the target digital word and encodes it into a much larger codeword space. It does this by interpreting the hash output as addresses in the larger codeword and places indelible marks - representing 1's – at the resulting address location. For a digital vector with n bits, represented by $x_{0,1,2..n}$ the hash function ($H(x)$) would produce marks at positions given by $H(x_0)$, $H(x_{0,1})$, $H(x_{0,1,2})$ and so on to $H(x_{0,1,2..n})$. This is shown schematically in [fig 1]. or additional message vectors the process is repeated, overlaying marks into the same codeword. Once all message vectors are included the codeword can be transmitted. The received codeword is decoded as follows: The receiver uses the hash function to determine the mark positions for $H(x_0=0)$ and $H(x_0=1)$ and then checks the received codeword to see if marks are present at either of these positions. If these marks are present they form the first live branches of a decoding tree. Decoding proceeds by calculating marks positions for $H(x_{0,1} = 00)$, $H(x_{0,1} =10)$, $H(x_{0,1} =01)$ and $H(x_{0,1} =11)$ (assuming both $H(x_0 =0)$ and $H(x_0 =1)$ are live). Where corresponding marks are found in the codeword branches are kept alive, where no marks are found no further possibilities are explored in subsequent rounds. This is repeated for bit position in the input vector. After the final round the message vectors remaining will contain the message vectors originally encoded. In the case where a large amount of noise is present, there may be additional false decodings, known as hallucinations. These hallucinations will obscure the genuine messages, but the original messages will always be present. Hallucinations can be killed off by adding additional fixed value checksum bits to the original messages such as $x_{n+1, n+2}=11$. This increases the number of decoding rounds and therefore the amount of processing required but is very effective at removing hallucinations. Burst errors, such as could be caused by signal blockage or fading, can appear as blocks of zeros in the codeword. These apparent 'gaps' in the codeword mark density can be identified when their size is statistically unlikely to appear by chance. If the decoding algorithm produces mark positions in the identified gaps, the algorithm can keep the branches alive. Because the data is distributed globally localised burst errors can be overcome by bridging across these gaps. Burst errors can be quite large – as much as 40% of the codeword – yet perfect decoding can be maintained. Even beyond the decoding schemes capacity to decode without errors all original message vectors are decoded but with some hallucinations present. The statistical significance of any gap is determined by the density of marks in the codeword. Thus, a minimum number of encoded message vectors are required to ensure the codeword is evenly populated with marks

In this paper 2 types of concurrent code are defined– open codes and closed codes, focussing predominantly on closed codes. Differences in computational load and signal to noise for the two types are examined. Synchronisation issues are then investigated and determine that closed concurrent codes have an inherent synchronisation capability. Comparisons are then made with CDMA to determine relative strengths and weaknesses regarding energy efficiency and robustness.

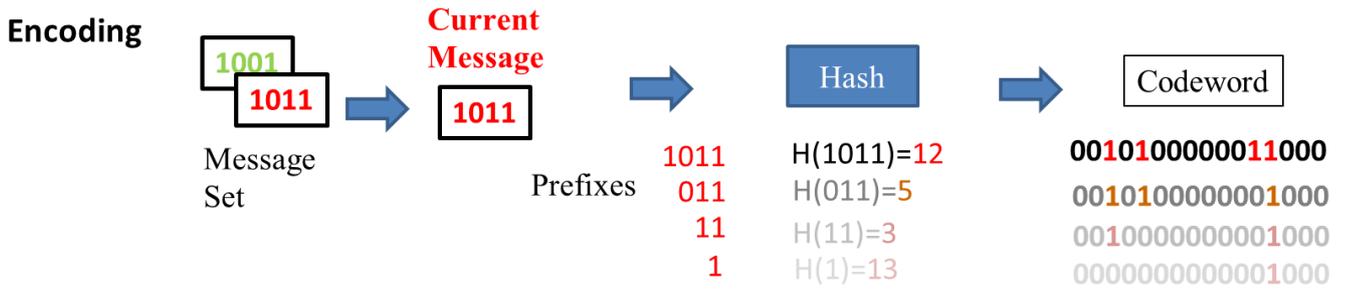

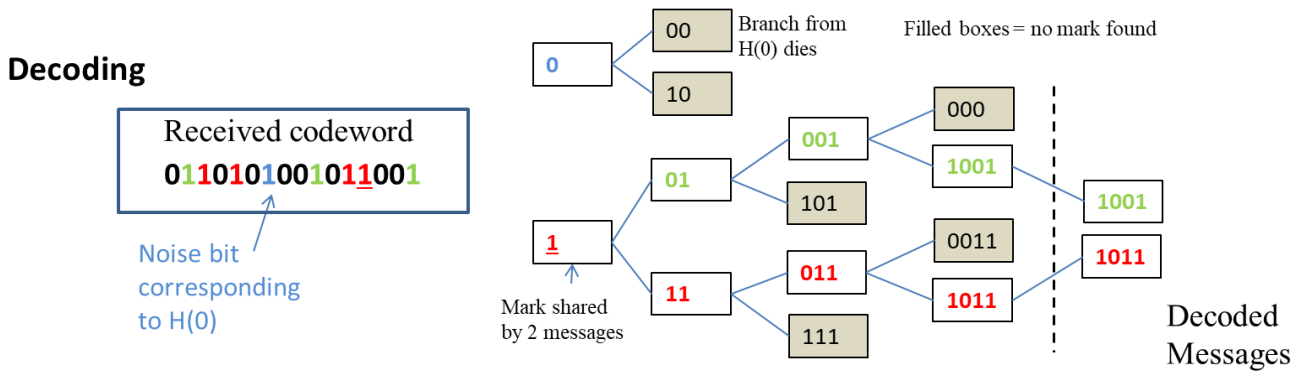

*Figure 1. A schematic representation of the concurrent code process for encoding (upper diagram) and the decoding tree (lower diagram). Input message prefixes are hashed to produce the addresses of marks in the codeword. The codeword is examined by comparing possible prefix hashes with the presence of marks in the codeword.*

## 2. Methods: Fundamental characteristics of concurrent codes

In this section is evidence of some characteristics of concurrent coding relating to the type of coding used, the relative efficiency in terms of transmitted energy and computational load, and the effects of noise.

### *2.1 Open and Closed Codes*

In the handful of publications to date regarding concurrent codes 2 regimes have been used all under the same headline of concurrent codes. In the earlier work large message vectors of several thousand bits were hashed into a codeword that could be millions of bits in length[3]. Other examples have used much smaller message vectors of 8 bits with codewords of a few thousand bits in length[6]. It is useful to define two types of code as follows:

**Closed Codes** – Where all possible message vectors can be uniquely encoded into the codeword. Thus $N$ bit messages, with $k$ checksum bits will all uniquely encode into a codeword of length $2^{N+k+1}$ bits.

**Open Codes** – where the number of possible message vectors is too high to be uniquely accommodated within the codeword. So a 1000 bit message vector (N+k) would require a codeword of $2^{1001} = 2 \times 10^{301}$ bits to be uniquely encoded which is clearly unfeasible.

The choice of open or closed code comes with different benefits and disadvantages. In both cases it is true that different messages that share common prefixes – such as 1001 and 1101 - will both share common marks in the codeword - so both message vectors produce marks at *H(1)* and *H(01)* and in total 6 marks are produced to represent 8 bits of information. The significance comes as more unique messages are added into the codeword. In a closed code a large fraction of the marks in the codeword are shared between message vectors as more prefixes are common and the number of marks increases logarithmically. This leads to a very efficient usage of the resources that represent marks. Open codes with long message vectors share very few bits and there is effectively a linear relationship between the number of message vector bits encoded and the number of marks. Thus closed codes can be significantly more efficient than open codes - and indeed than most other types of code – but here is the rub. In a closed code any given message vector can only be decoded once from the codeword because multiple inputs will give the same results in the codeword. This could be problematic when breaking down a data set into message vector units such as ASCII codes for text where repetition is likely. Thus, the input data will need to be conditioned to avoid repetition or be sourced such that repetition is unlikely. No such conditions apply to open codes where large message vectors can be sent. In theory hashing large messages into restricted codeword could lead to collisions and ambiguity but in practice large messages make this extremely unlikely [1],[3].

The use of open or closed codes will also affect the computational load required in the decoding stage. One helpful comparison here is to use a fixed codeword length and compare open and closed encoding for the same number of data bits. Lets take a block of *A* bits of information. This can be encoded in *m* message vectors of 8bit length. Including 2 checksum bits with each message requires a codeword of 2048 bits in length. The open coding uses a single message of *A* bits in length. We can calculate the number of marks produced *Z(m)* for the closed coding from the following:

$$Z(m) = Nm - m\log_2 m + \frac{3m}{2} \qquad (1)$$

Where *m* is number of messages being encoded and *N* is the message length. This is the equation presented in [6] with a modification of adding *3m/2* which provides a more accurate representation of the marks produced (although why this is so is not clear). In the decoding process every mark found in the message leads to 2 calls to the hash function. The number of marks for an open coding will be equal to the number of data bits, hence the number of hash function calls is simply *2A*. The number of hash function calls is related to the computational load of decoding. A plot of the hash function calls for open and closed codes is given in Figure 2. From this plot it is clear that closed coding is a much more efficient way of encoding when considering the marks produced and the computational load on decoding. From here on we will predominantly concentrate on closed codes.

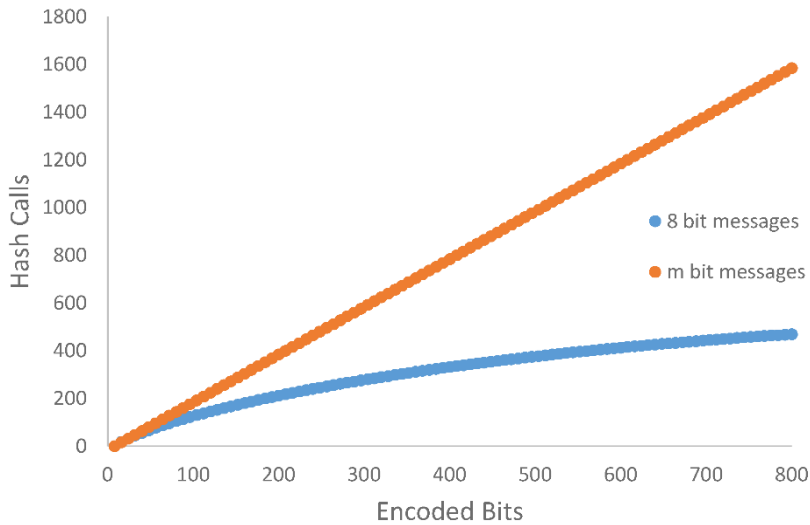

*Figure 2. The number of calls to the hash function in decoding for open coding (m bit message vectors) and closed coding (8 bit message vectors).*

### 2.2 Noise effects

The effect of noise in the codeword during decoding is to keep alive branches that would otherwise die thus increasing the number of calls to the hash function. At each decoding round each genuine mark will connect with a noise mark with probability $\mu$ which is the fraction of the codeword filled with noise. Thus the number of calls to the hash function *hc(m)* is given by:

$$hc(m) = 2Z(m) + \mu Z(m) \qquad (2)$$

The probability that a connected random mark finds another random mark to continue the branch goes as $\mu^2$ (and so on) which is small as decoding proceeds. There should then be a small second order term which we are ignoring to get an idea of broad characteristics.

Figure 3 shows a plot of the measured number of calls to the hash function vs number of encoded messages in the cases of µ=0 and µ=0.3 and good agreement is seen with the above equation. This was performed using 8 bit messages and a codeword of 2048 bits in length.

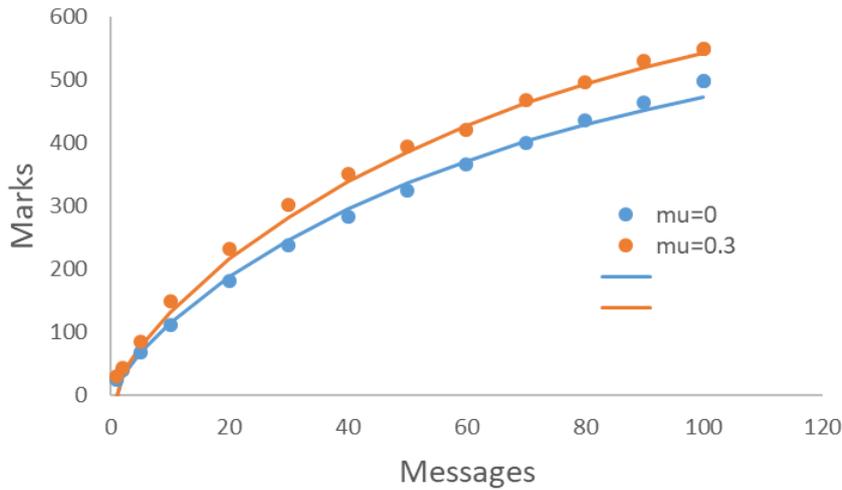

*Figure 3. Measured and modelled calls to the hash function vs number of encoded messages with noise fractions of 0.0 and 0.3.*

A concurrent codeword can have a variable number of message vectors encoded in it thus the signal to noise ratio of any particular codeword is dependent upon the number of messages and a constant noise level, although without decoding it is not possible to know in advance what this ratio is other than by assuming all marks are noise with just a single message encoded. The number of genuine marks is related to the number of messages via equation (1) and this is the signal. But a large number of messages fills up a lot of marks - does this reduce the noise component? Yes and no. Indelible marks mean noise and signal are indistinguishable and therefore can coexist. They are not additive but the relevant noise is that which does not include the signal marks. Thus the S:N is

$$S(m) = \frac{Z(m)}{\mu(2^{n+k+1} - Z(m))} \qquad [3]$$

where $\mu$ is the noise fraction. The maximum number of signal marks is $2^{n+1}$, $\mu$ cannot exceed 0.5 and $\mu$ should not exceed the hallucination threshold [6]. It is however convenient to plot the noise to signal ratio as shown in Figure 4.

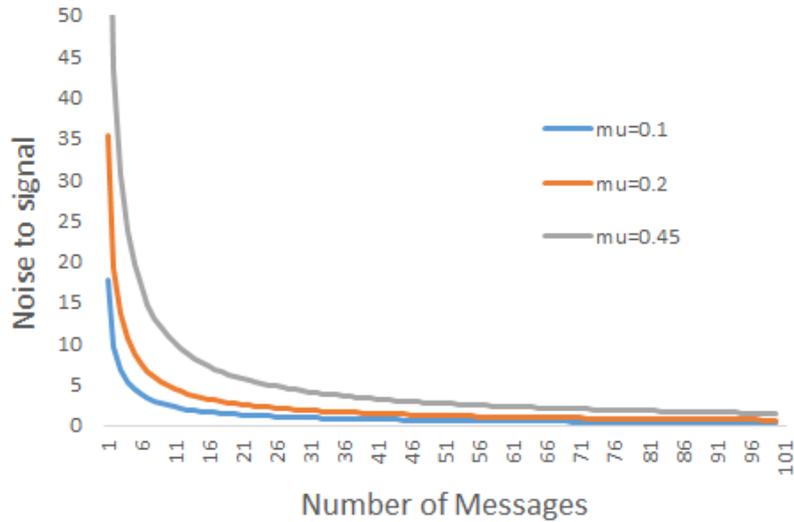

*Figure 4 The noise to signal ratio vs number of encoded message vectors. Plots for different noise levels are shown.*

The effect of the checksum bits is to kill hallucinations when noise is present. This was examined by adjusting the number of checksum bits and the required codeword length, then measuring the number of hallucinations generated as the noise fraction is increased. This can be seen in Figure 5. With no checksum bits and a 512 bit codeword, hallucinations appear immediately as noise is introduced. With 2 checksum bits hallucinations start to appear around 30% noise in a 2048 bit codeword. In this case the hash function used was a randomly generated hash-table generated to ensure no collisions. The position where hallucinations begin to appear has some dependence on the type of hash function used.

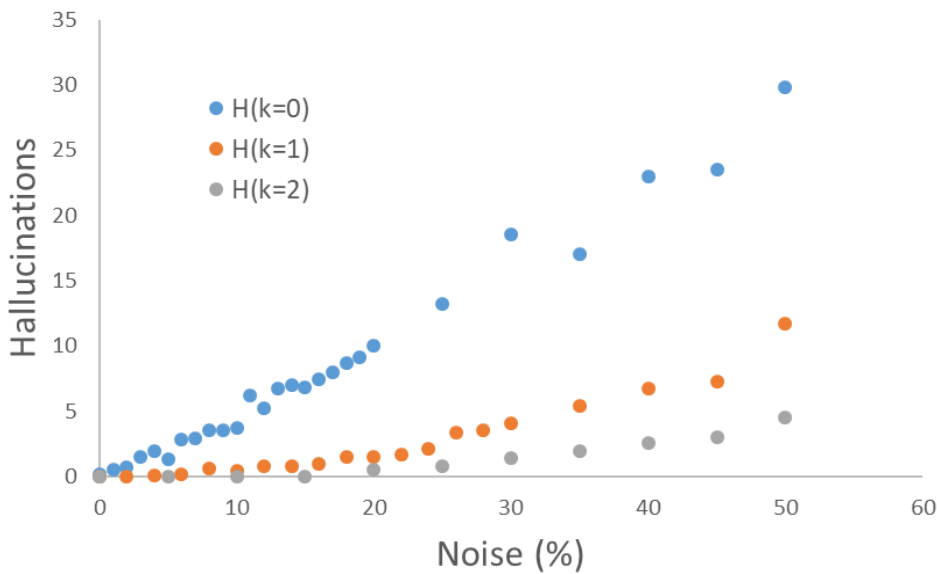

*Figure 5. Hallucinations generated vs the noise fraction. Plots are shown using different numbers of checksum bits and demonstrates how checksum bits kill hallucinations. Each point is the mean of 30 repeats.*

## 3. Results

### *3.1 Encoding with inherent synchronisation*

Concurrent coding uses indelible marks to represent a digital value of 1 and the absence of a mark represents 0. The first mark to be seen in a data stream is therefore very unlikely to be the first position in the codeword. In software testing we can define a definite zero position in a memory register, but for real signals the zero is an absence of signal. A synchronisation signal must then be used or incorporated into the data stream.

Ordinarily, without a separate specific synchronisation signal it would be impossible to reliably determine where a codeword started as there could be an indeterminate number of zeros (no signal) before the first mark. Thus synchronisation codes would be essential and synchronisation adds to the 'cost of the code' and reduces energy efficiency[19][20]. This would add to the overheads, reduce the code rate slightly and perhaps more importantly provide an attacker with the means to implement *smart* jamming. Disrupting a few bits in the synchronisation code – which must be easily identifiable - would render the code undecipherable. Whilst the concurrent coding is very resilient to noise and loss, the synchronisation code is not. This need for synchronisation was identified by Baird, Bahn and Collins[1] in their original work intended to produce an encoding mechanism robust against conventional jamming. This potential weakness can now be removed through the use of inherent synchronisation.

The concurrent coding principle encodes a digital word into a much larger codeword space by hashing incrementally increasing subsections (or prefixes) of the digital word to produce addresses in the larger codeword into which indelible marks are placed to represent 1's. All digital words that have a least significant bit (LSB) of 0 share the mark located at *H(0)*, and likewise all words with LSB of 1 share the marks at *H(1)*. This sharing of marks leads to efficient use of resources. Thus 50% of all possible words produce H(0) and 50% produce H(1). Whilst the position of these marks in the codeword is dependent upon the specifics of the hashing function, their existence is not. With even a handful of digital words encoded, marks in positions H(0) and H(1) will be present with high probability. Thus if the hashing function is known to the receiver, the detection of these marks can be used as a means of synchronisation. These marks are referred to as *primary* marks. The use of just 2 marks would be unreliable and a better system would be to make use of the next layer of hash encoding using the 6 marks generated from H(0), H(1), H(00), H(01), H(10) and H(11). Each of the second layer marks is produced with a 25% probability and are referred to as *secondary* marks. These 6 marks represent a unique pattern within the codeword determined by the hash function and are collectively referred to as the *principle* marks. Thus the receiver can generate this principle 6 mark pattern and correlate against the received codeword to determine the correct synchronisation. This technique has been demonstrated in an optical communication system using inherent synchronisation with multi-threaded encoding [21], but in this current work the details and features of inherent synchronisation are presented.

If we are to perform a correlation of the received codeword against the ideal principle mark pattern, we need to know how many of the principle marks positions might be filled in the random selection of message vectors included in the codeword. Consider the codeword tree. When a message is added to the codeword it must effectively occupy a unique position in the *Nth* decoding layer. This will determine which of all the prior branches are traversed in the decoding. We are interested in which of the secondary marks are used (primary marks are automatically used) and so we divide the final layer into quarters. The placing of the first message fills a primary and a secondary mark. The next message (assumed to be placed at random) will fill additional principle marks dependent on which quarter of the Nth layer is used, and the probability for this is simple to calculate. See diagram in for a schematic representation of principle marks. *P(y|x)* is defined to be the probability that *y* principle marks result from *x* initial marks after the addition of 1 message vector.

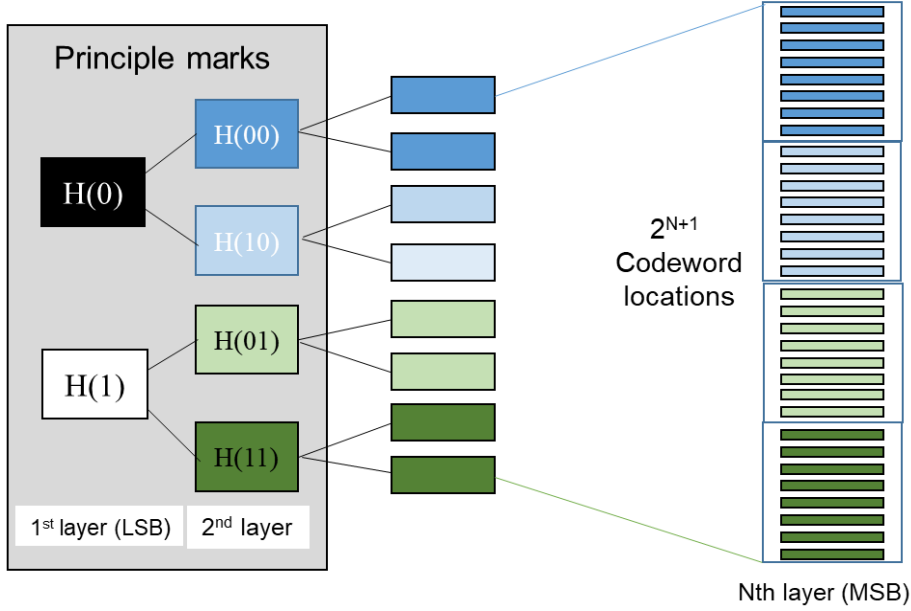

*Figure 6. A schematic diagram of the decoding tree showing the link between principle marks and final decoding layer*

The specific probability values are:

$$P(2|2) = 1/4 \quad P(3|2) = 1/4$$
$$P(4|2) = 1/2$$
$$P(3|3) = 1/2 \quad P(4|3) = 0$$
$$P(5|3) = 1/2$$
$$P(4|4) = 1/2 \quad P(5|4) = 1/2$$
$$P(5|5) = 3/4 \quad P(6|5) = 1/4$$
$$P(6|6) = 1 \tag{4}$$

The probability of A marks being filled after *i* words are added, $P(A)_i$ is:

$$P(2)_i = P(2)_{i-1} P(2|2)$$

$$P(3)_i = P(2)_{i-1} P(3|2) + P(3)_{i-1} P(3|3)$$

$$P(4)_i = P(2)_{i-1} P(4|2) + P(3)_{i-1} P(4|3) + P(4)_{i-1} P(4|4)$$

$$P(5)_i = P(3)_{i-1} P(5|3) + P(4)_{i-1} P(5|4) + P(5)_{i-1} P(5|5)$$

$$P(6P(6)_i = P(5)_{i-1} P(6|5) + P(6)_{i-1} P(6|6) \tag{5}$$

With the initial condition that $P(2)_1 = 1$ and all other initial states are zero.
From these probabilities the average number of principle marks present after m words are added is simply:

$$zp_m = \sum_{a=2}^{6} a\, P(a)_m \tag{6}$$

This is also the value of the correlation coefficient to be expected which is telling us how many of the principle marks are present.

Using 8 bit message vectors with 2 check sum bits and a $2^{11}$ bit codeword space, a quantity of random message vectors were hashed into the codeword space. Correlation against the principle mark pattern was then performed and repeated to generate a mean correlation value. This was repeated for various numbers of message vectors and plotted in Figure 7. Also plotted is the expected correlation calculated with the above method and good agreement is seen.

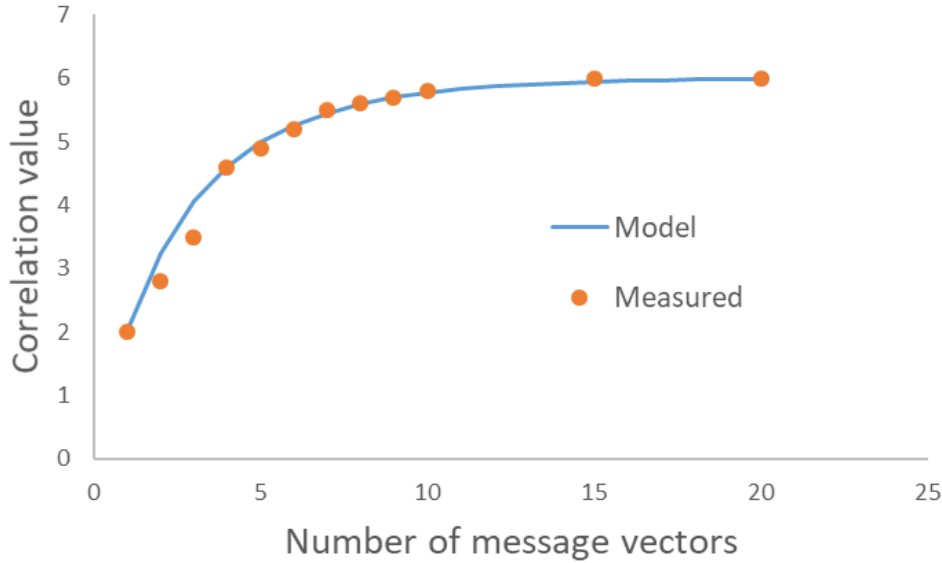

*Figure 7. Correlation values for a given number of random message vectors, both measured and modelled. Each data point represent 30 repeats of the process.*

In concurrent coding there is no fixed number of message vectors that are contained within a codeword, this number can vary and may contain only a single message vector if that is all that is required. There are however constraints on concurrent coding performance. One constraint is that the ability to correct for burst errors requires reliable identification of gaps (a significant run of contiguous zeros) in the codeword and this is achieved with a minimum number of message vectors present (assuming no noise). The same is true for synchronisation. Reliable synchronisation will require the correlation to exceed a threshold value with little likelihood of random correlations occurring at this value. A look at the plot suggests that with 6 messages included we should expect a correlation value of around 5, with 10 messages approaching the maximum value of 6.

The number of q-fold correlations that arise from a random noise fraction of $\mu$, in a codeword space of $2^N$ is:

$$f(q) = 2^N \mu^q \quad (7)$$

Choosing *q=5*, with *N=11*, the number of 5 fold correlations varies with noise as shown in Figure 8. The large number of correlations at high noise do not corrupt the synchronisation, they simply offer a large number of potential alternatives that must be tried until the correct synchronisation is found.

We can determine an acceptable noise level by requiring the number of random q-fold correlations to be less than 1, say *f(5)=0.1*, thus the acceptable noise level would be

$$\mu_{q=5} = \sqrt[5]{\frac{0.1}{2^N}} = 0.137 \quad (8)$$

If we set the acceptable number of random correlations to be higher, say f(5)=1 then the acceptable noise level, would be $n_{q=5}=0.216$.

The existence of inherent synchronisation means that a single, isolated codeword can be transmitted and reliably detected, synchronised and decoded with no special accompanying signatures. All that is required of the decoder is a knowledge of the data rate and a knowledge of which hash function to use. Indeed this has been tried and observed to be the case and will be reported separately.

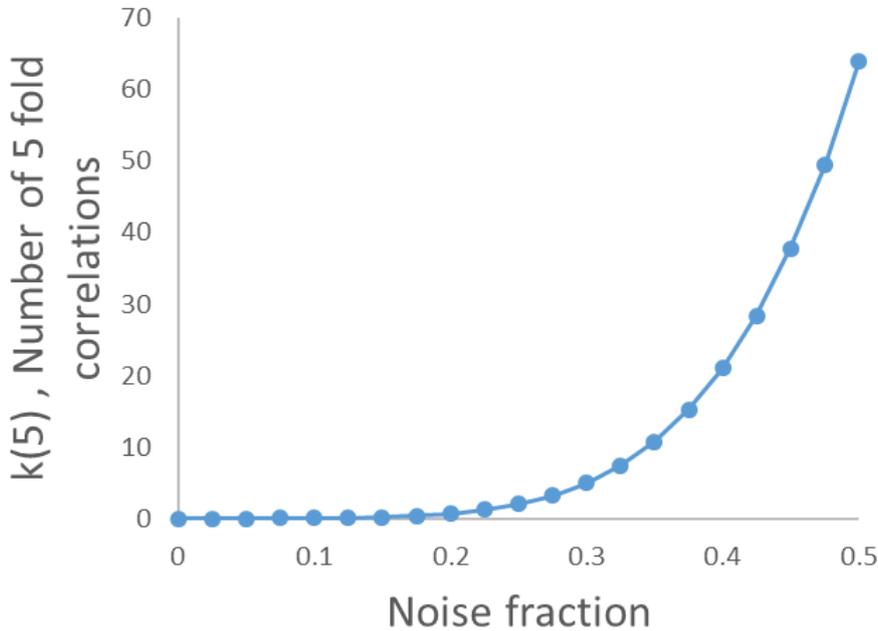

*Figure 8. A plot of the number of 5 fold correlations against the noise level in the codeword.*

The main implication of inherent synchronisation is that the apparent limit on the acceptable noise level is lower than the noise threshold level where hallucinations begin to appear in the decoding. This might suggest that inherently synchronising is the limiting factor of the performance of concurrent codes, but this is not necessarily so. Firstly, inherent synchronisation need not be used and conventional synchronisation could be used, with the understanding that smart jamming becomes a potential weakness. The process outlined here works with a single codeword in isolation. By combining the correlations of multiple codewords in succession the correct synchronisation will become apparent at higher noise levels but at the expense of increased latency. Inherent synchronisation requires no additional data content to be added to the transmission, ensuring the minimum transmitted energy, but will require a correlation across the whole codeword and may therefore add to the computational load. A synchronisation pattern could potentially be identified quickly with a smaller number of correlations.

Burst errors are an issue for synchronisation. A gap that removes marks from a portion of the codeword can remove principle marks with the effect of reducing the correlation value attained. This could see the correlation fall below a synchronisation threshold. However as gaps need to be identified for the purposes of burst error correction, this same principle could be used to reduce the synchronisation threshold. At low noise levels this is still unlikely to prevent the identification of correct synchronisation.

### *3.2 Comparison between Concurrent coding and CDMA*
Concurrent coding was originally conceived as a method for providing protection against jamming, an alternative to spread spectrum techniques. Concurrent coding can achieve jamming resistance without the need for a shared secret key as is required in a code division multiple access (CDMA) scheme. Comparison of the behaviour of these encoding methods is therefore appropriate.

CDMA takes each bit of data to be transmitted and performs an XOR with a binary spreading code of length *L* which can be just a few bits or up to 256 bits. This spreads the original information across a wider bandwidth and offers a processing gain related to the length of the spreading sequence. Data from other users of the channel who use a different spreading code, can be XORed together. Decoding proceeds by XORing the received codeword with a particular spreading code appropriate to the user. This produces a binary word of length L which will be all the same value if no noise is present. When noise alters decoded values the largest number of 0's or 1's determines the final decoded data bit.

When comparing a concurrent code with a CDMA approach it is essential to understand that the two encoding systems are performing different actions and producing outputs in very different ways. How then do we perform a like for like comparison?

We start with the assumption that we a have a certain amount of data that we wish to transmit, with the knowledge that the transmission channel may be noisy and difficult. We then require some knowledge of how concurrent coding and CDMA would behave. Concurrent coding of the closed form takes a message of *N* bits, appends *k* checksum bits and then hashes them into a codeword space that is $2^{N+k+1}$ bits in size. We start by choosing the same codeword size for both concurrent code and CDMA, so that both techniques are using the same amount of temporal resource and bandwidth. In this comparison we will assume a simple approach where message vectors are 8 bits long with 2 checksum bits. Thus the required codeword length is 2048 bits. The total number of bits to be encoded using CDMA is dependent upon the error correction scheme being used. In this case an (8,4) Hamming code was used. The interleaving spacing is determined by the encoded block size (8). The total number of messages m determines the length of the spreading code (chip code) that can be used from *codeword length/number of encoded bits*. The spreading code size is the level of processing gain $G_{CDMA}$ obtained by encoding 1 bit across many

$$G_{CDMA} = \frac{2^{N+k+1}}{mNf} \qquad (9)$$

where *f* is the expansion factor of the error correcting code being used.

From this we can see that shorter chip codes arise from more messages being sent and hence there is reduced protection against jamming for more data being sent in a given codeword.

The equivalent gain for concurrent code can be considered as the ratio of codeword size to message size:

$$G_{CC} = \frac{2^{N+k+1}}{N+k} \qquad (10)$$

Notice this does not change with the number of encoded messages.

### 3.3 Random noise

A number of 8 bit messages were chosen at random. These messages were then Hamming encoded, interleaved and spread to produce a final codeword. Messages were split into 4 bit blocks before Hamming encoding and then cross interleaved. This codeword was then corrupted by adding noise marks randomly and/or removing contiguous regions to represent burst errors. The corrupted codeword was then decoded and the level of error recorded. The same approach was taken with data encoded with concurrent coding before corruption. The effect of the corruption with CDMA leads to the correct number of messages but some incorrectly decoded. With concurrent coding the corruption leads to additional messages being falsely decoded. To compare the two approaches we consider the decoded error fraction, that is, the fraction of decoded messages that are incorrectly decoded. The effects of noise on the decoding is shown in Figure 9 where the noise fraction is the fraction of the codeword that is affected by random noise marks in addition to the message data.

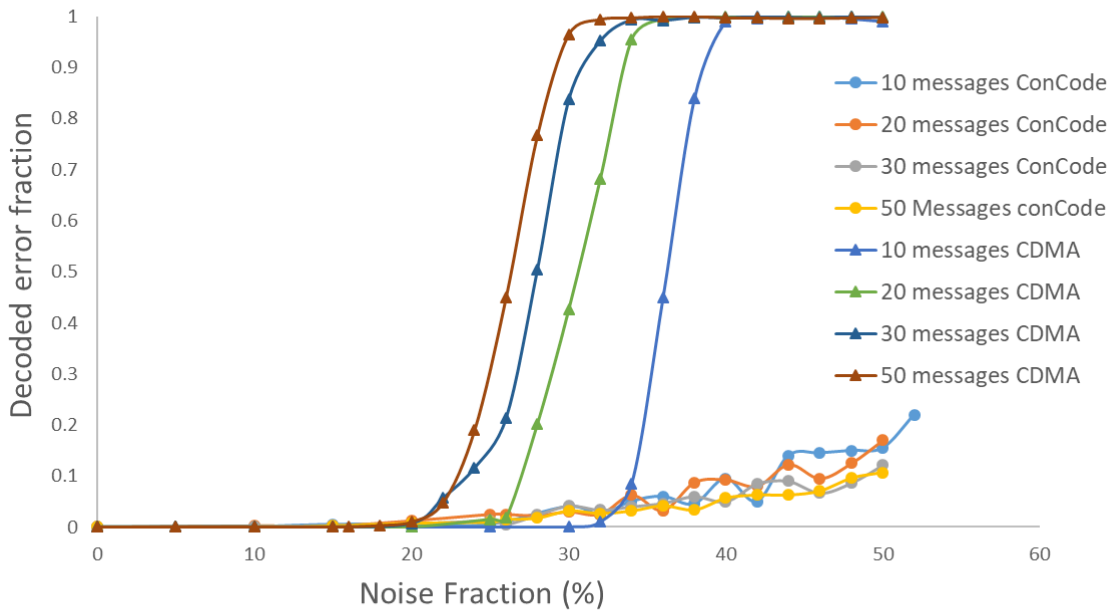

*Figure 9. Decoded error fractions for CDMA and concurrent coding where a given number of messages are encoded into a fixed size codeword and noise is added before decoding. Each point is the mean of 50 repeats of the process.*

The effect of the reduced processing gain for CDMA is evident in earlier onset of errors as more messages are incorporated. Once the error correcting capacity is breached the errors increase rapidly with additional noise levels until all encoded messages are corrupted. In contrast concurrent coding still correctly decodes the original messages but hallucinations are produced. Some hallucinations start to be generated at noise levels where the CDMA correctly decodes when the number of messages is low and the processing gain is high. At noise levels below 20% (S:N =10*8/2048 = 0.03) both approaches perform well. At higher noise levels CDMA performance is dependent on number of messages but above 40% noise will deliver no information. Concurrent coding will deliver correctly decoded information but it would need to be winnowed out from the hallucinations somehow.

### *3.4 Burst errors*
The effect of introducing burst errors is shown in Figure 10. Contiguous sections of data were set to zero before decoding and the size of the zeroed section is given as a fraction of the codeword size and called the gap fraction. Errors start to appear with CDMA encoding as the gap fraction exceeds 12%. This corresponds to twice the interleaving spacing and is the limit beyond which the error correction scheme can correct for errors. The concurrent coding performs well with just a few hallucinations occasionally appearing at 20% and then steadily increasing. At a gap fraction of 40% CDMA produces no correctly decoded information and concurrent coding obscures correct decoding with hallucinations. The negative dip at low gap fractions for concurrent codes arises from gaps being smaller than a threshold gap size and thus not identified as gaps with no subsequent correction implemented in the decoding tree.

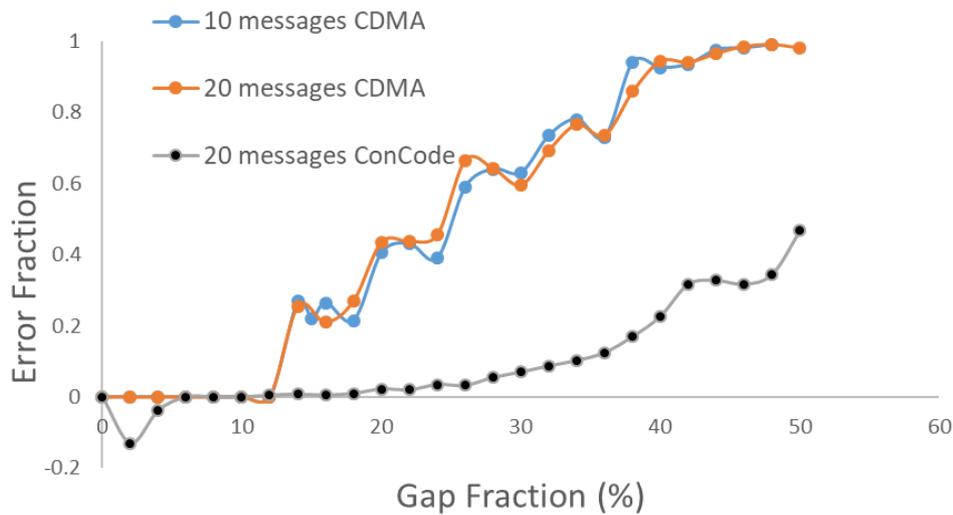

*Figure 10 decoded error fractions for CDMA and concurrent coding as contiguous gaps of data are introduced into the codeword. Each point is the mean of 50 repeats.*

Spread spectrum encoding plays no part in helping to correct for burst errors as this is entirely dependent upon the interleaving and error correction process.

To achieve comparable burst error performance would require larger interleaving spacings, which would require small block sizes for the original data. The error correction scheme for this may be just a bit repetition. Alternatively larger interleave spacings could be achieved by using a longer codeword and spacing out the data further. This is not like for like comparison unless we are considering what is necessary to achieve specific performance against burst errors. In which case concurrent codes would seem superior to spread spectrum techniques.

### *3.5 Transmission Efficiency*

We have assumed that both encoding schemes are using a similar modulation, such as an On-Off Keying, as this is an appropriate method for producing indelible marks. Such a system only uses energy when it transmits a mark, thus a determination of the number of 1's being generated is a measure of the efficiency of the transmission scheme. Concurrent coding makes efficient use of marks in 2 ways. Firstly marks are only produced for original data bits, so a small amount of data produces a small number of marks. Secondly concurrent coding produces degenerate marks that can represent many common sub-sequences simultaneously. In contrast CDMA uses spreading codes which have equal number of 0's and 1's and thus 50% of the codeword consists of a 1, irrespective of the number of encoded bits. Thus CDMA always transmits more energy than a concurrent code as can be seen in Figure 11 where the number of transmitted marks is plotted against the number of original data bits that they represent. The concurrent code is always more efficient. The maximum number of CDMA encoded messages is 128, for which the number of marks produced is 1024, compared to 448 for concurrent coding. This represents at a minimum a 228% relative efficiency improvement with values over 1000% achievable at low data content.

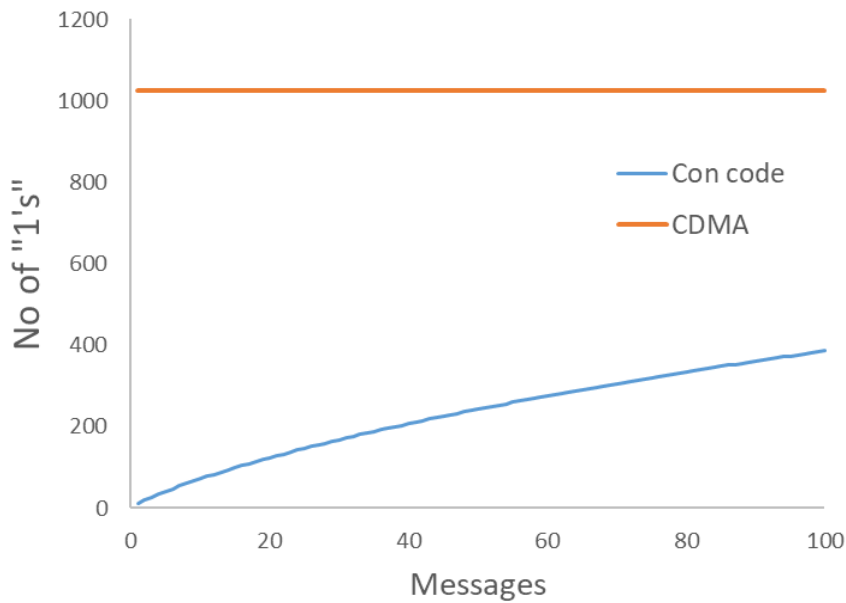

*Figure 11. Transmission efficiency in terms of marks produced.*

There are clearly situations where a constant output power level is desirable, such as is the case with 3G mobile phone systems. However there are other scenarios where reducing the required output power would be desirable, either for conserving energy, minimising interference with other systems or ensuring a low probability of intercept. In these cases concurrent coding would certainly be advantageous.

### 3.6 Computational load

The computation involved in CDMA is well understood. The error correcting code – in this case a Hamming code – and interleaving are straightforward and quick to compute, as is the XOR function with the spreading code. The same process is run in decoding making CDMA symmetrical for transmitter and receiver. Concurrent coding however is not symmetrical. The computational load is governed by the number of calls to the hash function and the computational complexity of the hash function. The complexity of the hash function can be radically different depending on its nature and hence we can only think in generic terms regarding the hash function calls. For transmission the hash function is called once for every data bit and additionally once to hash all the checksum bits with each message. Thus the computation increases linearly with the data being encoded. In decoding the hash function is called twice for every true message mark present. Thus in an ideal case with no noise present the number of calls to the hash function goes as *2Z(m)* where *Z(m)* is the number of marks generated by *m* messages given in equation (1). With noise present the number of hash calls is given by equation (2).

The plots in Figure 12 show the number of hash function calls for the encoding and decoding phases with various levels of noise having been added increasing the decoding load. The most interesting aspect of this can be seen at high relative data content. A codeword containing a small amount of data (small number of messages) requires twice as many hash calls for decoding compared to encoding. However for larger amount of encoded data fewer hash calls are required for decoding compared to encoding. This asymmetry, where decoding is computationally easier than encoding, could be a feature unique to concurrent coding.

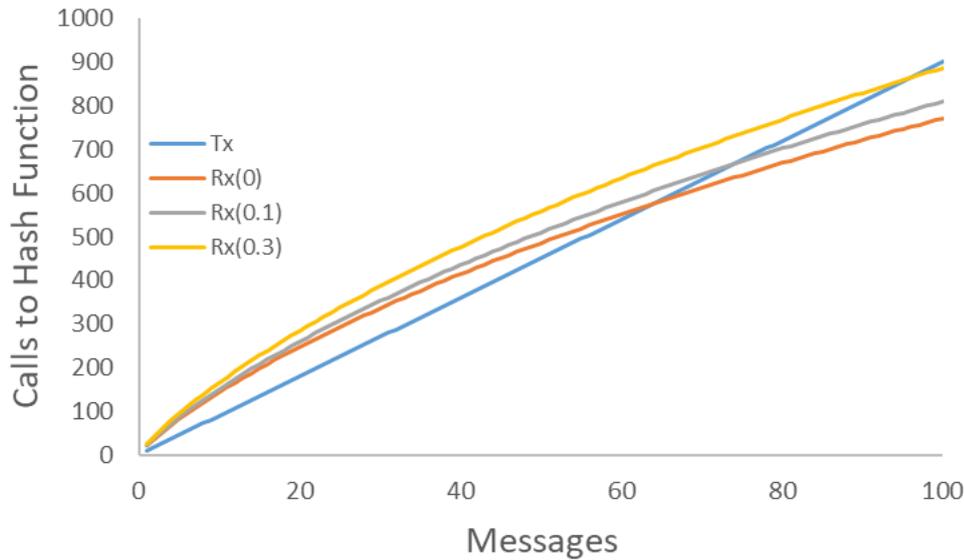

*Figure 12. Number of calls to the hash function for the transmission and decoding stages with various levels of noise in codeword.*

Whilst the absolute load is dependent upon the hash function complexity, it is intuitive to see that CDMA is much simpler and faster to implement than concurrent coding with fewer calls to the error correcting function.

## 4. Conclusions

Concurrent coding is presented here as an encoding method that could offer an alternative way of encoding to protect against noise, burst errors and jamming. Concurrent coding operates by encoding information from included message vectors globally across the whole codeword transmission. In contrast most error correction encoding techniques encode information locally and broaden the spread of information through the use of systematic interleaving. Concurrent coding is a single step to achieve all these characteristics and is simple to understand, especially in comparison to cyclic encoding.

This work has highlighted the significant features attributable to concurrent coding that could define its suitability for particular applications. In particular when using a closed concurrent code the transmitted energy can be significantly reduced in comparison with other encoding techniques. This could find application in energy efficient systems or could be valuable where a low probability of intercept is required. In addition to this you also get robustness against noise, burst errors and jamming. The behaviour against large burst errors is particularly impressive where concurrent coding provides significantly more robustness than does interleaving for a given codeword length.

The need for synchronisation between transmitter and receiver is critical to any communication system. Concurrent coding had been originally designed to provide protection against brute force jamming, but the need for synchronisation provided a weakness against smart jamming that would target a few bits within a synchronisation code. Because in concurrent coding a lack of signal represents a zero, the first mark(1) to be seen is unlikely to be the start of the codeword therefore synchronisation is needed to infer where the codeword starts. In this work we have shown that no synchronisation code is required because concurrent code contains an inherent synchronisation code that is built up statistically with the addition of a few message vectors to the codeword. Synchronisation is achieved by correlation against a set of six principle marks that represent the initial branches of the decoding tree. It is therefore possible to send an isolated codeword comprising a set of marks amidst a sea of zeros, and successfully synchronise and decode. The correlation process can be susceptible to noise issues and there needs to be a minimum number of messages included to ensure successful correlation. This minimum number is between 6-8 message vectors and is a little higher than the number needed to ensure identification of data gaps used in the correction of burst errors (typically 5). The effect of noise upon correlation is to confuse the correlation process by providing many potential positions of high correlation. The threshold noise level is lower than the level at which hallucinations start to be produced. It is therefore clear that the use of inherent synchronisation is a limiting step in the use of concurrent coding.

In choosing an encoding scheme to provide protection against noise, burst errors and jamming it would be normal to reach for a CDMA scheme. In a like-for-like comparison with CDMA concurrent coding performs very well, with a slower degradation against noise, better performance against burst errors and significantly better transmission efficiency. It should be pointed out that the CDMA scheme used is only as good as the underlying error correction codes that it is built upon. The CDMA spreading process adds processing gain to the underlying forward error correction scheme to help against noise and jamming but the burst error protection is generally just interleaving. Concurrent coding may require more computational resources, although this is highly dependent upon the nature of the hashing function being used. Concurrent coding can display the unusual characteristic of requiring less computational effort to decode than to encode when sufficient message vectors are incorporated into the codeword.

Concurrent codes therefore present an attractive alternative to established and conventional encoding techniques and would be especially attractive to techniques requiring low power transmission. Concurrent coding is also very simple to understand which is an advantage over some other methods such as cyclic encoding, or a combination of methods such as error correction, interleaving and spread spectrum. The biggest issue with using a closed concurrent code is the inability to represent the same message twice as any message can only be decoded once. This can be addressed by combining data and positional values into messages and the use of a restricted codebook. Techniques involving multiple simultaneous hashing functions, where the same data can be represented with different functions, are currently under investigation.

*Acknowledgements*

The author would like to acknowledge useful discussions with Paul Brittan.

## 5. References

[1] Baird LC.III, Bahn W L. & Collins MD. 'Jam-Resistant Communication Without Shared Secrets Through the Use of Concurrent Codes', Technical Report, U. S. Air Force Academy;2007.USAFA-TR-2007-01 .

[2] Baird LC.III Bahn, W L, Collins, MD Carlisle M C. & Butler S. 'Keyless jam resistance', Proceedings of the 8th Annual IEEE SMC Information Assurance Workshop (IAW), Orlando, Florida, June 20-22,;2007 pp 143-150. doi: 10.1109/IAW.2007.381926

[3] Bahn W L. Baird L C. III & Collins, M D. 'Jam resistant communications without shared secrets', Proceedings of the 3rd International Conference on Information Warfare and Security (ICIW08), Omaha, Nebraska,;2008 April 24-25.

[4] Baird LC.III, Carlisle M & Bahn WL 'Unkeyed Jam Resistance 300 Times Faster: The Inchworm Hash', MILCOM 2010 - Military Communications Conference, San Jose, CA, Oct;2010

[5] Baird LC III, Schweitzer D, Bahn WL & Sambasivam S. 'A Novel Visual Cryptography Coding System for Jam Resistant Communications', Journal of Issues in Informing Science and Information Technology, 7,2010, pp 495-507.

[6] Benton DM (2016) 'Concurrent Codes: A Holographic-Type Encoding Robust against Noise and Loss'. PLoS ONE 11(3): e0150280. doi:10.1371/journal.pone.0150280

[7] Martin Bossert. 'Channel Coding for Telecommunications' (1st ed.). *John Wiley & Sons*, Inc., New York, NY, USA; 1999.

[8] Glover I & Grant PM 'Digital Communications', *Pearson Education.* ;2010.

[9] Sklar B, 'Digital Communications: Fundamentals and Applications', Second Edition *Prentice-Hall*,;2001, ISBN 0-13-084788-7.

[10] Berrou C, Glavieux A & Thitimajshima P 'Near Shannon limit error-correcting coding and decoding: Turbo-codes. 1'. In Communications, 1993. ICC 93. Geneva. Technical Program, Conference Record, IEEE International Conference on , Vol. 2, pp. 1064-1070. IEEE;1993

[11] Viterbi, A J 'Error bounds for convolutional codes and an asymptotically optimum decoding algorithm'. IEEE Transactions on Information Theory 13 (2): pp 260–269.;1967doi:10.1109/TIT.1967.1054010

[12] Lin, Shu, and Daniel Costello. 'Error Control Coding: Fundamentals and Applications.' *Prentice Hall*;1983

[13] Clark Jr, George C., and J. Bibb Cain. 'Error-correction coding for digital communications.' *Springer Science &*

*Business Media*; 2013.
[14] Fire, P. 'A class of multiple-error-correcting binary codes for non-independent errors', Sylvania Rept. RSL-E-2, Sylvania Reconnaissance Systems Laboratory, New York ;1959
[15] http://www.usna.edu/Users/math/wdj/files/documents/reed-sol.htm, accessed February 2016.
[16] Baird,LC, Carlisle MC, Bahn, WL & Smith E. 'The Glowworm hash: Increased speed and security for BBC unkeyed jam resistance'. In military communications conference, 2012-milcom (2012, October). (pp. 1-6). IEEE
[17] Baird, L. C., & Parks, B. (2015). 'Exhaustive attack analysis of BBC with glowworm for unkeyed jam resistance.' Proceedings - IEEE Military Communications Conference MILCOM, 2015-Decem,300–305. http://doi.org/10.1109/MILCOM.2015.7357459
[18] Hamid Hanifi, Leemon Baird, and Ramakrishna Thurimella. 2015. 'A new algorithm for unkeyed jam resistance.' In Proceedings of the 8th International Conference on Security of Information and Networks (SIN '15). ACM, New York, NY, USA, 210-216. DOI: http://dx.doi.org/10.1145/2799979.2800008
[19] Li, G. Y., Xu, Z., Xiong, C., Yang, C., Zhang, S., Chen, Y., & Xu, S. (2011). 'Energy-efficient wireless communications: Tutorial, survey, and open issues.' IEEE Wireless Communications, 18(6), 28–35. http://doi.org/10.1109/MWC.2011.6108331
[20] Chandar, V., Tchamkerten, A., & Tse, D. (2013). 'Asynchronous capacity per unit cost.' IEEE Transactions on Information Theory, 59(3), 1213–1226. http://doi.org/10.1109/TIT.2012.2236914
[21] Benton, D., & St. John Brittan, P. (2017). A new encoding scheme for visible light communications with applications to mobile connections. Advanced Free-Space Optical Communication Techniques and Applications III, 11. https://doi.org/10.1117/12.2277321.